\documentclass[12pt,preprint]{aastex}

\newcommand{\mdot}{\mbox{$\dot M$}{}}

\begin{document}
\title{Diffuse X-rays from the Inner 3 Parsecs of the Galaxy}

\author{Gabriel Rockefeller$^{1,2}$, Christopher L. Fryer$^{1,2}$,
  Fulvio Melia$^{1,3,4}$, and Michael S.  Warren$^{2}$}
\altaffiltext{1}{Physics Department, The University of Arizona, Tucson, AZ
  85721}
\altaffiltext{2}{Theoretical Division, LANL, Los Alamos, NM 87545}
\altaffiltext{3}{Steward Observatory, The University of Arizona, Tucson, AZ
  85721}
\altaffiltext{4}{Sir Thomas Lyle Fellow and Miegunyah Fellow.
}

\begin{abstract}  
  Recent observations with the {\it Chandra} X-ray Observatory have
  provided us with the capability to discriminate point sources, such
  as the supermassive black hole Sgr A*, from the diffuse emission
  within the inner $10^{\prime\prime}$ of the Galaxy.  The hot plasma
  producing the diffuse X-radiation, estimated at $\approx 7.6\times
  10^{31}$ ergs s$^{-1}$ arcsec$^{-2}$ in the 2--10 keV band, has a
  RMS electron density $\approx 26$ cm$^{-3}$ and a temperature
  $kT\approx 1.3$ keV, with a total inferred mass of $\approx
  0.1\;M_\odot$.  At least some of this gas must be injected into the
  ISM via stellar winds.  In the most recent census, about 25 bright,
  young stars have been identified as the dominant sources of the
  overall mass efflux from the Galactic center.  In this paper, we use
  detailed 3-dimensional SPH simulations to study the wind-wind
  interactions occurring in the inner 3 parsecs of the Galaxy, with a
  goal of understanding what fraction, if any, of the diffuse X-ray
  flux measured by {\it Chandra} results from the ensuing shock
  heating of the ambient medium.  We conclude that this process alone
  can account for the entire X-ray flux observed by {\it Chandra} in
  the inner $10^{\prime\prime}$ of the Galaxy.  Understanding the
  X-ray morphology of the environment surrounding Sgr A* will
  ultimately provide us with a greater precision in modeling the
  accretion of gas onto this object, which appears to be relatively
  underluminous compared to its brethren in the nuclei of other
  galaxies.
\end{abstract}


\section{Introduction}

The inner few parsecs of the Galaxy contain several observed
components that coexist within the central deep gravitational
potential well.  The pointlike, nonthermal radio source Sgr A* appears
to be the radiative manifestation of a $\sim 3$ million M$_\odot$
concentration of dark matter within $\sim 0.015$ pc of the nucleus.
This concentration, probably a central black hole, is surrounded by a
cluster of evolved and young stars, a molecular dusty ring, ionized
gas streamers, diffuse hot gas and Sgr A East, a shell-like,
nonthermal radio source believed to be the remnant (SNR) of a
supernova explosion at the center of the Galaxy some $10,000$ years
ago \citep[see][]{MF01}.

Observations at infrared and, particularly, radio wavelengths have
been very effective in unraveling the complex behavior of the central
components through their mutual interactions.  But only through recent
observations with the {\it Chandra} X-ray Observatory have we been
able to study the X-ray emission from Sgr A* and its surroundings with
sub-arcsecond resolution and a broad X-ray band imaging detector.
These capabilities have provided us with the means to discriminate
point sources, such as Sgr A*, from the diffuse emission due to hot
plasma in the ambient medium.

\citet{BEA03} report that the {\it Chandra}/ACIS-I 0.5--7 keV
image of the inner $17^{\prime}\times 17^{\prime}$ region of the
galaxy contains remarkable structure; its detail is sufficient to
allow comparisons with features seen in the radio and IR wavebands.
The diffuse X-ray emission is strongest in the center of Sgr A East.
Based on these X-ray observations, \citet{MEA01} conclude that
Sgr A East is a rare type of metal-rich, ``mixed-morphology" SNR that
was produced by the Type II explosion of a 13--20 $M_\odot$
progenitor.  The X-ray emission from Sgr A East appears to be
concentrated within the central 2--3 pc of the $6\times9$ pc radio
shell and is offset by about $2$ pc from Sgr A* itself. The thermal
plasma at the center of Sgr A East has a temperature $kT\approx 2$
keV, and appears to have strongly enhanced metal abundances with
elemental stratification.

However, this is not the sole component of hot gas surrounding Sgr A*.
{\it Chandra} has also detected X-ray emission extending perpendicular
to the Galactic plane in both directions through the position of Sgr
A*, which would seem to indicate the presence of a hot, ``bipolar''
outflow from the nucleus.  And an additional component of X-ray
emitting gas appears to have been found within the molecular ring,
much closer to Sgr A*. Indeed, the western boundary of the brightest
diffuse X-ray emission within $2$ pc of the Galactic center coincides
very tightly with the shape of the Western Arc of the thermal radio
source Sgr A West.  Given that the Western Arc is believed to be the
ionized inner edge of the molecular ring, this coincident morphology
suggests that the brightest X-ray emitting plasma is being confined by
the western side of the gaseous torus. In contrast, the emission
continues rather smoothly into the heart of Sgr A East toward the
eastern side.
 
A detailed fit of the emission from the hot gas within
$10^{\prime\prime}$ of Sgr A* shows that it is clearly not an
extension of the X-ray emitting plasma at the center of Sgr A East.
The inferred 2--10 keV flux and luminosity of the local diffuse
emission are $(1.9\pm 0.1)\times 10^{-15}$ ergs cm$^{-2}$ s$^{-1}$
arcsec$^{-2}$ and $(7.6^{+2.6}_{-1.9})\times 10^{31}$ ergs s$^{-1}$
arcsec$^{-2}$, respectively.  Based on the parameters of the best-fit
model, it is estimated that the local, hot diffuse plasma has a RMS
electron density $\langle n_e^2\rangle^{1/2}\approx 26$ cm$^{-3}$.
Assuming that this plasma has a filling factor of unity and is fully
ionized with twice solar abundances, its total inferred mass is
$M_{\hbox{\rm local}}\approx 0.1\;M_\odot$.  The local plasma around
Sgr A* appears to be distinctly cooler than the Sgr A East gas, with a
temperature $kT=1.3^{+0.2}_{-0.1}$ keV.  By comparison, the best-fit
model for Sgr A East indicates that the plasma there has about 4 times
solar abundances, and twice this temperature (see above).

At least some of this local, hot plasma must be injected into the ISM
via stellar winds.  There is ample observational evidence in this
region for the existence of rather strong outflows in and around the
nucleus. Measurements of high velocities associated with IR sources in
Sgr A West \citep{K91} and in IRS 16 \citep{G91}, the $H_2$ emission
in the molecular ring and from molecular gas being shocked by a
nuclear mass outflow (\citealp{G86}; but see \citealp{jack93} for the
potential importance of UV photodissociation in promoting this $H_2$
emission), broad Br$\alpha$, Br$\gamma$ and He I emission lines from
the vicinity of IRS 16 \citep{HKS82,AHH90,G91}, and radio continuum
observations of IRS 7 \citep{YM91}, provide clear evidence of a
hypersonic wind, with a velocity $v_w \sim$ 500--1,000 km~s$^{-1}$, a
number density $n_w\sim10^{3\mbox{--}4}$ cm$^{-3}$, and a total mass
loss rate $\dot M_w\sim$ 3--4 $\times10^{-3}\; M_\odot$~yr$^{-1}$,
pervading the inner parsec of the Galaxy.

\citet{C92} discussed the properties of stellar winds in the Galactic
center and their ability to produce X-rays, under the assumption that
the colliding winds create a uniform medium with a correspondingly
uniform mass and power input within the central 0.8~pc of the Galaxy.
While the assumption of uniformity may be an over-simplification,
steady-state models yield central conditions similar to those
calculated in our simulations and (apparently) observed in nature.

The possibility that wind-wind collisions in the dense stellar core of
the Galaxy might produce an observable X-ray signature was considered
more recently in the context of transient phenomena by \citet{OGU97}.
These authors argued that the strong winds from OB stars or WR stars
would produce shocks that heat the gas up to a temperature $T\sim
10^7$ K.  Instabilities in the overall outflow could therefore produce
transient high temperature regions, in which the X-ray luminosity
could be as high as $10^{33}$--$10^{35}$ ergs s$^{-1}$.

A preliminary, low-resolution 3-dimensional hydrodynamical simulation
by \citet{CM99} subsequently revealed that although the tesselated
pattern of wind-wind collision shocks can shift on a local dynamical
time scale, the overall concentration of shocks in the central region
remains rather steady.  This is hardly surprising, given that 25 or so
dominant wind sources are distributed rather evenly about the center.

In this paper, we study in detail the emission characteristics of
shocked gas produced in wind-wind collisions within 3 parsecs of Sgr
A*.  We present the results of comprehensive, high-resolution
numerical simulations of the wind-wind interactions, using the latest
suite of stellar wind sources with their currently inferred wind
velocities and outflow rates.  We use their projected positions and
vary the radial coordinates to determine the effect of this
uncertainty on the X-ray emissivity.  In addition, it is clear now
from the {\it Chandra} image that the stellar winds are also
interacting directly with the molecular ring, and so we incorporate
the latter into the set of boundary conditions for our simulations.
The predicted X-ray luminosity and spatial profile may then be
compared with the {\it Chandra} data.  We describe our numerical
technique, including the description of this molecular ring and the
characteristics of the wind sources in \S\ 2 of the paper. In \S\ 3 we
present our results, and discuss their implications for the conditions
in the Galactic center in \S\ 4.

\section{The Physical Setup}

Our calculations use the 3-dimensional smoothed particle hydrodynamics
(SPH) code discussed in \citet{FW02}, and \cite{War03}.  The
Lagrangean nature of SPH allows us to concentrate spatial resolution
near shocks and capture the complex time-dependent structure of the
shocks in the Galactic center.

We assume that the gas behaves as an ideal gas, using a gamma-law
($\gamma = 5/3$) equation of state.  Since the mass of gas in the
simulation at any time is many orders of magnitude less than the mass
of the black hole and its halo of stars, we calculate gravitational
effects using only the gravitational potential of the assumed central
$2.6\times 10^{6}\;M_\odot$ point mass.

Each of our two calculations uses slightly different initial
conditions; details of each calculation are summarized in
Table~\ref{sims}.  The initial conditions for each simulation assume
that the Galactic center has been cleared of all mass (perhaps by the
supernova explosion that produced Sgr A East).  Massive stars then
inject mass into the volume of solution via winds, gradually
increasing the mass of gas in the Galactic center.  The number of
particles in each simulation initially grows rapidly, but reaches a
steady state ($\sim 7$ million particles) when the addition of
particles from wind sources is compensated by the particle loss as
particles flow out of the computational domain or onto the central
black hole.  The particle masses vary from $9.4\times
10^{-7}\;M_\odot$ to $1.0\times 10^{-5}\;M_\odot$.

In addition, we place a torus of molecular material in a circumnuclear
disk surrounding the inner 1.2 pc of the Galaxy.  The boundary
conditions (outer and central black hole), the circumnuclear disk, and
the stellar wind sources are the only modifications made to the basic
SPH code of \citet{War03} to run these simulations, and we discuss
these modifications in more detail in the next three subsections.

\subsection{Boundary Conditions}

The boundary conditions are the major particle sinks in the
simulation.  We assume that once these particles have achieved escape
velocity beyond a certain distance, nothing confines the material
flowing out of the Galactic center.  The computational domain is a
cube approximately $6$ pc on a side, centered on the black hole.  To
simulate ``flow-out'' conditions, particles passing through this outer
boundary are removed from the calculation.

In addition, particles passing through an inner spherical boundary
with a radius of $1.9\times 10^{17}$~cm (equivalent to $2.5\times
10^5$~R$_s$ or $\sim 0.7$~Bondi-Hoyle capture radii for Sgr A*)
centered on the black hole are also removed.  This assumes that the
accretion onto Sgr A* continues roughly at free-fall below this point.
Such a simplified assumption only provides a limit on the X-ray
luminosity of the point-source of Sgr A*, but it does not affect the
diffuse X-ray emission in the Galactic center that is the central
topic of this paper.

\subsection{Circumnuclear Disk}

HCN synthesis data, taken at $5^{\prime\prime}-10^{\prime\prime}$
spatial and $4$ km s$^{-1}$ spectral resolution, reveal a highly
inclined, clumpy ring of molecular gas surrounding the ionized central
$\sim 2$ pc of the Galaxy \citep{G87}.  This ring seems to be the
inner edge of a thin molecular structure extending out to $\sim 5-7$
pc from the center.  Within this cavity, and orbiting about Sgr A*, is
the huge H II region known as Sgr A West comprised of a 3-armed
mini-spiral \citep{LC83,L91}.  The molecular and ionized gas in the
central cavity appear to be coupled, at least along the so-called
western arc (one of the spiral arms), which appears to be the ionized
inner surface of the molecular ring.  It has been suggested that the
northern and eastern arms may be streamers of ionized gas falling
toward Sgr A* from other portions of the molecular ring
\citep[see][and references cited therein]{MF01}.

Since 1987, observations of various tracers of molecular gas (notably
H$_2$, CO, CS, O I, HCN, NH$_3$) have somewhat refined our view of the
circumnuclear disk (or CND, as it is sometimes called), leaving us
with the following set of defining characteristics: a total mass of
$\sim 3\times 10^4\;M_\odot$, a very clumpy distribution with an
estimated volume filling factor of $\sim 0.01$, a typical clump mass
of $\sim 30\;M_\odot$, size $\sim 0.1$ pc, and temperature $\sim 100$
K \citep[see][]{G86,S86,G87,D89,S90,jack93,M93}.  As many as $500$
clumps fill its overall structure.  Its height at the inner edge is
$\sim 0.5$ pc, and it flares outward with distance from the center.

The {\it Chandra} $0.5-7$ keV image of the central $1.3^\prime\times
1.5^\prime$ region of the Galaxy (see Fig.~4 in \citealp{BEA03}),
overlaid on the VLA 6-cm contours of Sgr A* and Sgr A West, shows that
the western boundary of the brightest diffuse X-ray emission coincides
very closely with the shape of the western arc, adding credence to the
notion that the latter tracks the inner edge of the CND if one adopts
the view that the X-ray glowing plasma is itself being confined by the
western side of the molecular ring.

The CND therefore appears to be an essential element in the (partial)
confinement of the hot gas, and we incorporate it into our simulations
using a simple approach that nonetheless retains the CND's important
features.  To keep the volume of solution tractable, we do not model
the entire $\sim 14$-pc disk, but rather introduce an
azimuthally-symmetric torus with inner radius $1.2$ pc
\citep[see][]{G87} and thickness $1$ pc.  The latter is a fair
representation of the CND's observed height at this radius. This
structure does not, of course, account for the CND's outer regions,
but as we shall see, the hot, X-ray emitting gas does not penetrate
far into the molecular boundary, so the outer CND is not directly
relevant to this simulation.  We assume that this torus has a mass of
$\sim 10^4\; M_\odot$, and is composed of 200 spherical clumps, each
of mass $50\;M_\odot$.  We also incline it by $60^\circ$ from the
plane of the sky, orienting its principal axis (in projection) along
the Galactic plane.

\subsection{Wind Sources}

\begin{figure}
\epsscale{1.00}
\plotone{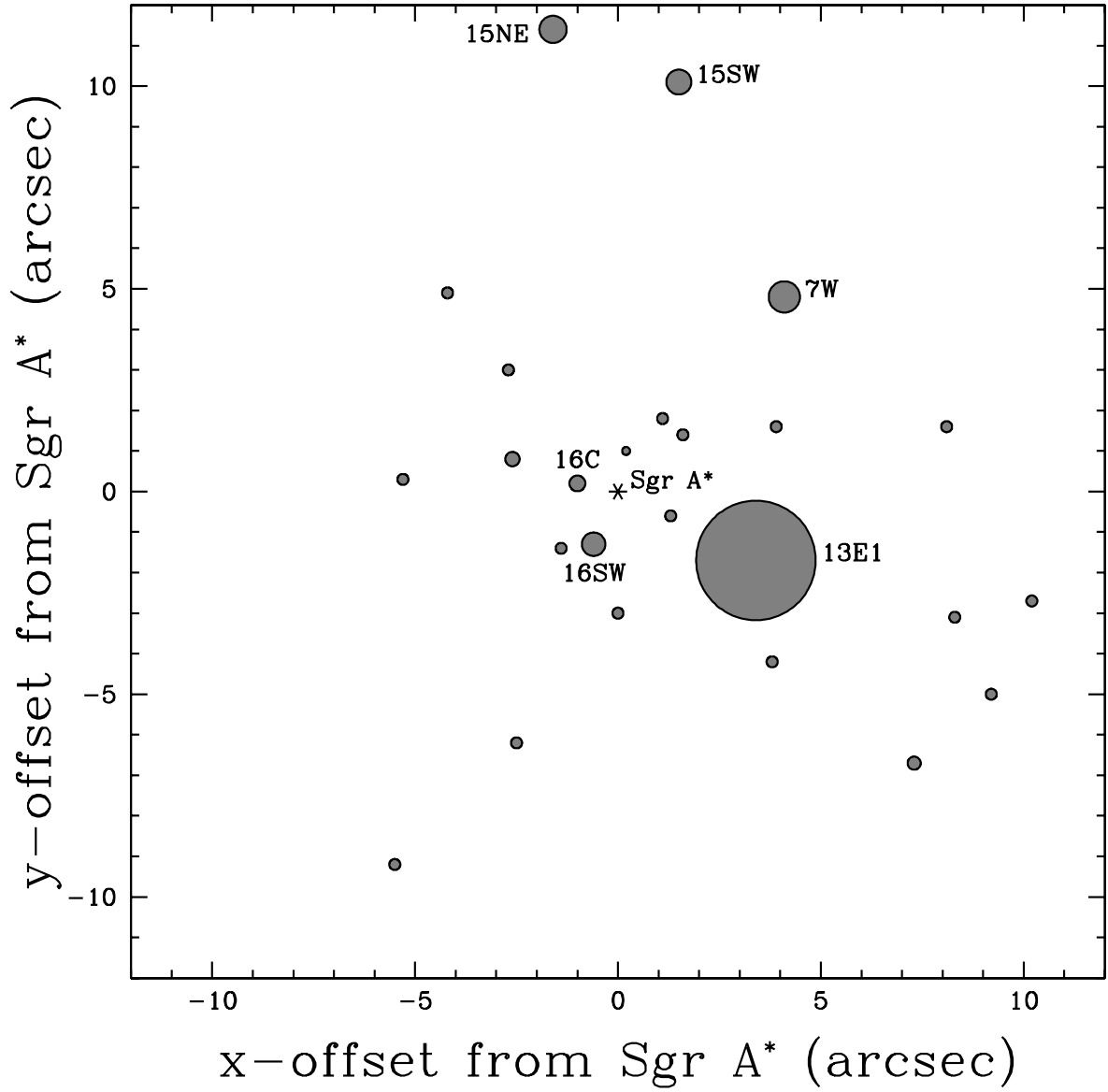}
\caption{Location of the 25 wind-producing stars used in the
  simulations reported here, relative to the position of Sgr A*
  indicated by the * symbol. The radius of each circle corresponds (on
  a linear scale) to that star's mass loss rate.  Setting the scale is
  13E1, with $\dot M=7.9\times 10^{-4}\;M_\odot$ yr$^{-1}$.}
\label{fig-winds}
\end{figure}

It is well known by now that the Galactic center wind is unlikely to
be uniform, since many stars contribute to the mass ejection
\citep[see, e.g.,][]{CM99}.  For these calculations, we assume that
the early-type stars enclosed (in projection) within the Western Arc,
the Northern Arm, and the Bar produce the observed wind.  Thus far, 25
such stars have been identified \citep{genz96}, though the stellar
wind characteristics of only 8 have been determined from their He I
line emission \citep{N97}.  Figure~\ref{fig-winds} shows the positions
(relative to Sgr A*; $1^{\prime\prime}\approx 0.04$~pc at the Galactic
center) of these wind sources; the size of the circle marking each
position corresponds to the relative mass loss rate (on a linear
scale) for that star.  IRS 13E1 and IRS 7W seem to dominate the mass
outflow with their high wind velocity ($\sim$~1,000~km~s$^{-1}$) and a
mass loss rate of more than $2\times10^{-4}\;M_\odot$ yr$^{-1}$ each.

Wind sources are modeled as literal sources of SPH particles.  New
particles are added in shells around each wind source as the
simulation progresses and the existing particles move outward from
each source.  Using wind velocities inferred from observations
\citep{N97}, shells around each wind source are added at a rate such
that the spacing between shells is approximately equal to the spacing
between particles within each incoming shell.  Masses of the incoming
particles are chosen to match the inferred mass loss rate for each
source.  The stars without any observed He I line emission are
assigned a wind velocity of 750~km~s$^{-1}$ and an equal mass loss
rate chosen such that the total mass ejected by the stars used here is
equal to $3\times10^{-3}\;M_\odot$ yr$^{-1}$, the overall mass outflow
inferred for the Galactic center region \citep[see, e.g.][]{Me94}.
The initial temperature of the stellar winds is not well known; for
simplicity, we assume here that all the winds are Mach 30 at their
point of ejection from the stellar surface. This corresponds to a gas
temperature of $\sim 10^{4}$ K.  We note, however, that our results
are insensitive to this value, since the temperature of the shocked
gas is set primarily by the kinetic energy flux carried into the
collision by the winds.  For the calculations reported here, the
sources are assumed to be stationary over the duration of the
simulation.

An additional uncertainty is the location in $z$ (i.e., along the line
of sight) of the wind sources.  To test the sensitivity of our results
to the choice of $z$, we carry out two simulations with different
coordinate assignments, and compare the overall shocked gas
configurations and their X-ray emissivities.  For the first choice of
$z$ positions---listed as $\rm z_{1}$ in Table~\ref{srcs} and used in
Simulation 1---we determine this third spatial coordinate randomly,
subject to the condition that the overall distribution in this
direction matches that in $x$ and $y$.  With this proviso, all these
early-type stars are located within the central few parsecs
surrounding Sgr A*.  For the second choice of positions---listed as
$\rm z_{2}$ and used in Simulation 2---we attempt to calculate a
reasonable upper limit on the luminosity produced in the central
parsec by moving all the sources closer together in $z$, toward the
$z=0$ plane.

\section{The Gas Profile and Spectrum}

\begin{figure}
\epsscale{1.00}
\plotone{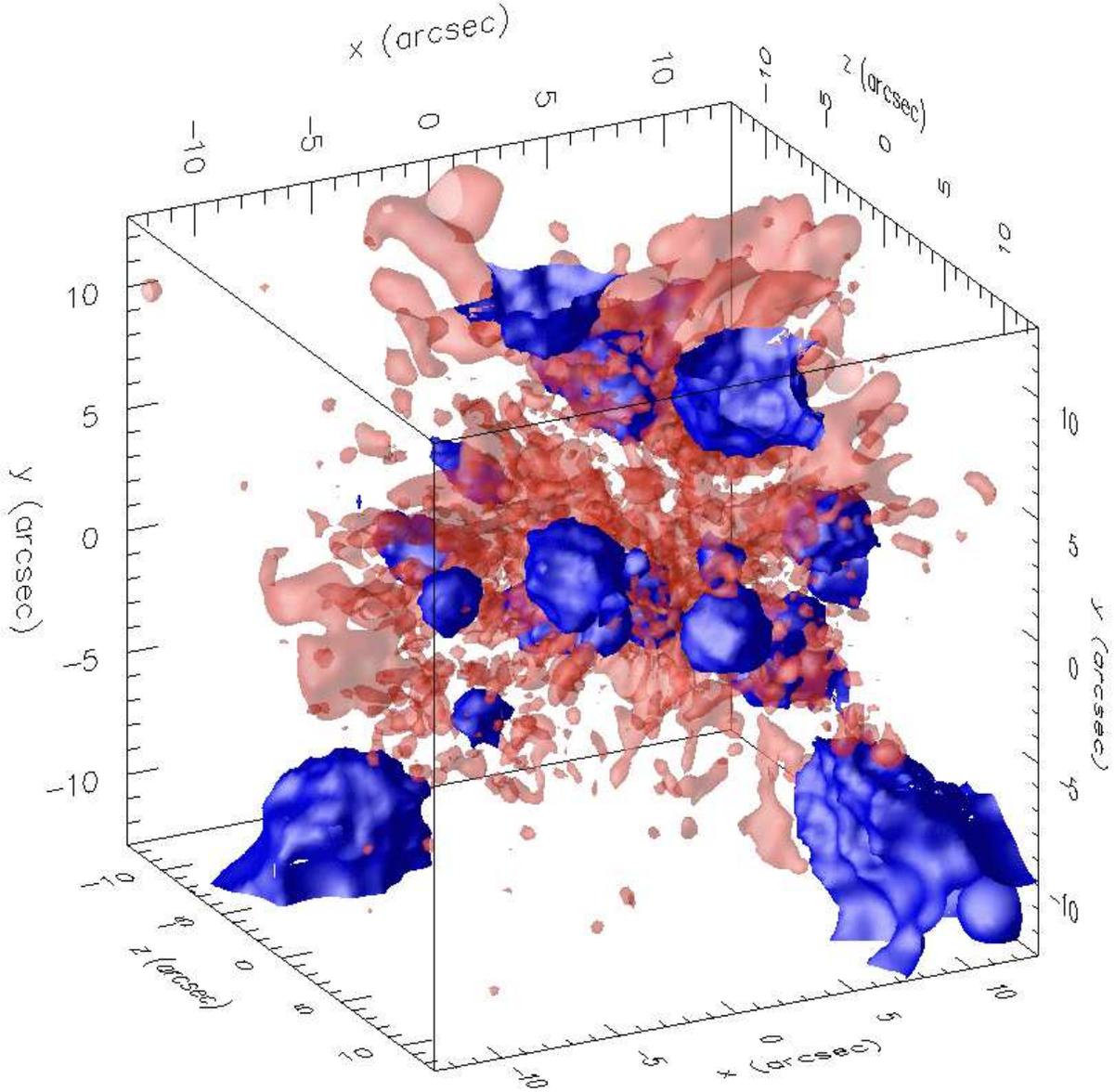}
\caption{Isosurfaces of specific internal energy from Simulation 1,
  $10,000$ years after the winds were turned on.  The blue surfaces
  correspond to a specific internal energy of $2.5\times
  10^{12}$~ergs~g$^{-1}$; the red surfaces correspond to $3.8\times
  10^{15}$~ergs~g$^{-1}$.}
\label{fig_3D}
\end{figure}

Wind-wind collisions in the central parsec create a complex
configuration of shocks that efficiently converts the kinetic energy
of the winds into internal energy of the gas.  Figure~\ref{fig_3D}
shows isosurfaces of specific internal energy in the central cubic
parsec of Simulation 1, $\sim~10,000$ years after the beginning of the
calculation.  The blue surfaces indicate regions of gas with low
specific internal energy, which lie near the wind sources; the red
surfaces mark regions of high specific internal energy, where gas has
passed through multiple shocks.  26\% of the total energy in the
central parsec has been converted to internal energy via multiple
shocks: the total kinetic energy of material in the central parsec is
$7.7\times 10^{48}$~ergs, while the total internal energy is
$2.7\times 10^{48}$~ergs.

In order to calculate the observed continuum spectrum, we assume that
the observer is positioned along the negative $z$-axis at infinity and
we sum the emission from all of the winds---and their
shocks---produced by the 25 stars in Table~\ref{srcs}. For the
conditions we encounter here, scattering is negligible and the optical
depth is always less than unity.  For these temperatures and
densities, the dominant components of the continuum emissivity are
electron-ion ($\epsilon_{ei}$) and electron-electron ($\epsilon_{ee}$)
bremsstrahlung.

\begin{figure}
\epsscale{1.00} 
\plotone{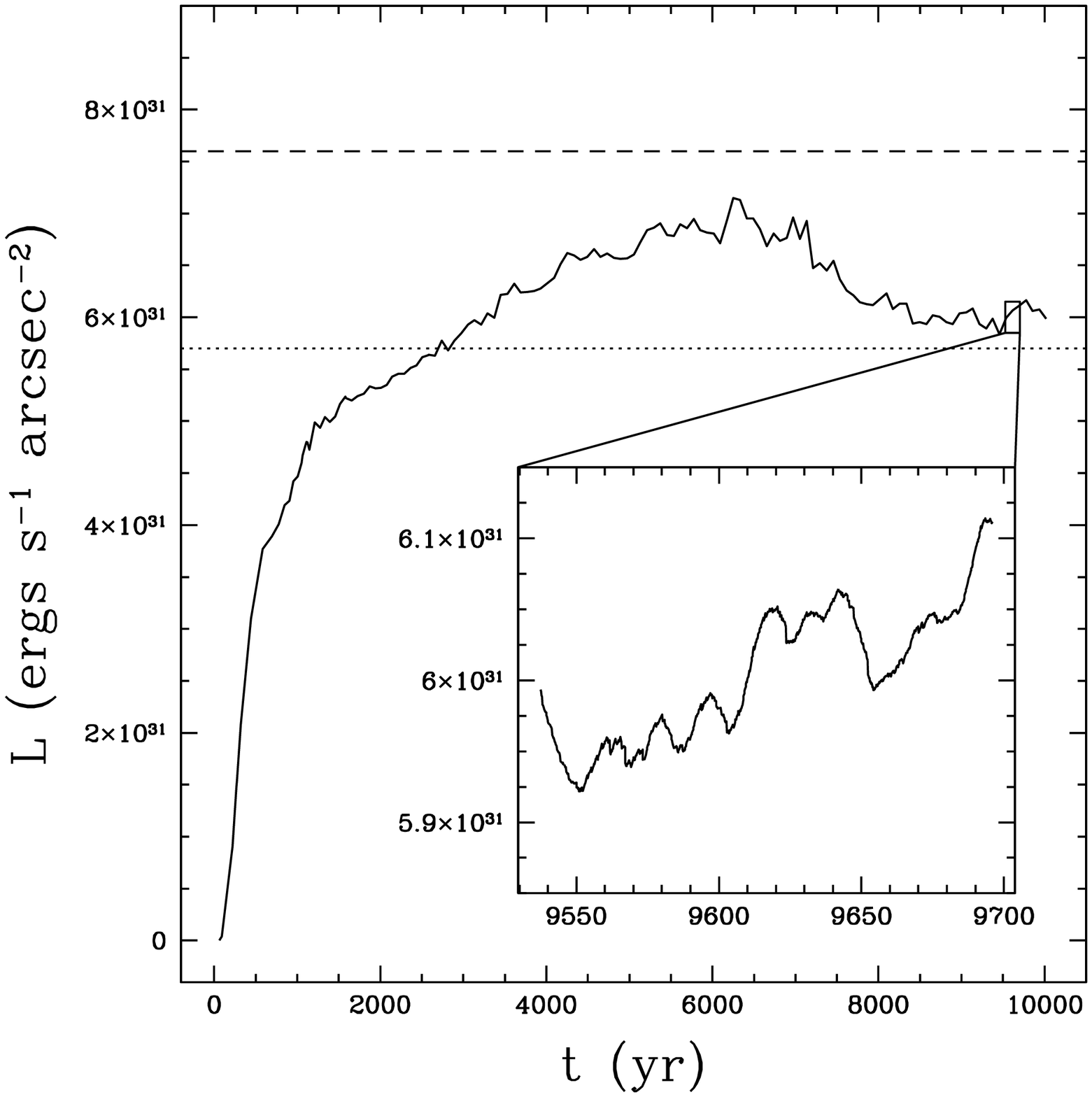}
\caption{2--10 keV X-ray luminosity per arcsec$^{2}$ from the central
  10$^{\prime\prime}$ versus time for Simulation 1.  The large plot
  shows the variation in luminosity over the entire calculation, while
  the inset plot shows variation on a timescale of $\sim$~10~years.
  The winds fill the volume of solution after $\sim$~4000 years.  The
  dashed line indicates the luminosity measured by {\it Chandra},
  while the dotted line indicates the value of the lower error bar on
  that measurement.  The upper error bar, at $1.02\times 10^{32}$ ergs
  s$^{-1}$ arcsec$^{-2}$, is not visible on this graph.}
\label{fig-lvst}
\end{figure}

\begin{figure}
\epsscale{1.00} 
\plotone{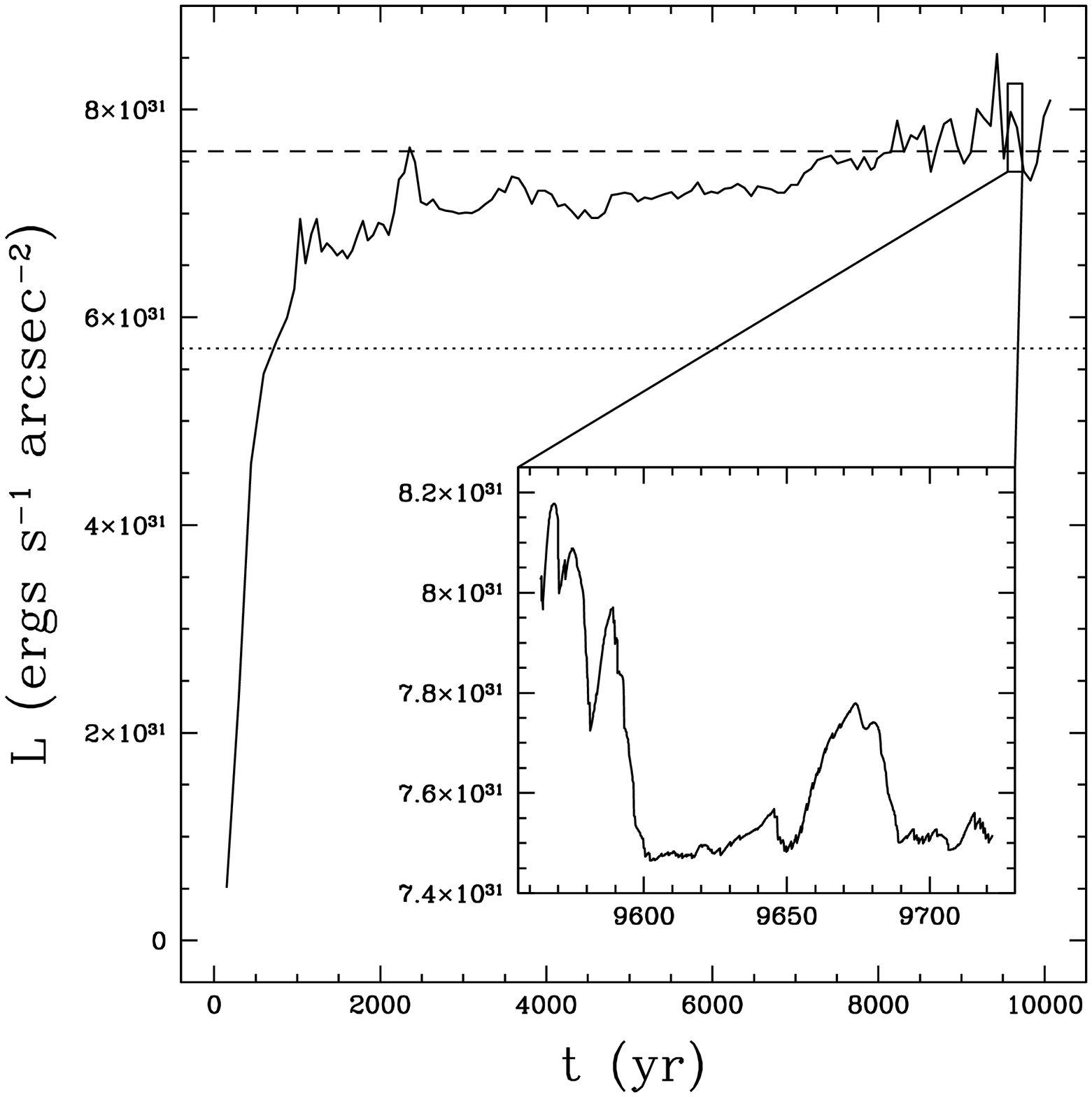}
\caption{2--10 keV X-ray luminosity per arcsec$^{2}$ from the central
  10$^{\prime\prime}$ versus time for Simulation 2.  The large plot
  shows the variation in luminosity over the entire calculation, while
  the inset plot shows variation on a timescale of $\sim$~10~years.
  The winds fill the volume of solution after $\sim$~4000 years.  The
  dashed line indicates the luminosity measured by {\it Chandra},
  while the dotted line indicates the value of the lower error bar on
  that measurement.  The upper error bar, at $1.02\times 10^{32}$ ergs
  s$^{-1}$ arcsec$^{-2}$, is not visible on this graph.}
\label{fig-lvst-lowm}
\end{figure}

Figures \ref{fig-lvst} and \ref{fig-lvst-lowm} show the 2--10 keV
luminosity per arcsec$^{2}$ from the central 10$^{\prime\prime}$ of
Simulations 1 and 2, respectively, versus time since the beginning of
each calculation.  The winds fill the volume of solution after
$\sim$~4000 years, which coincides with the point in each figure where
the luminosity stops rising rapidly and begins varying by only
$\sim$~5\% around a steady value.  The inset plot in each figure shows
variation of the luminosity on a timescale of $\sim$~10~years.  Data
points in the insets are plotted for each of 990 time steps, where the
size of each time step is 0.16 years.  Variation between consecutive
data points is due primarily to numerical noise, but variations on
timescales of several years indicate noticeable changes in the
temperature and density of the gas in the central 10$^{\prime\prime}$.

The range of values of the average luminosity per arcsec$^{2}$
measured in the central 10$^{\prime\prime}$ of our simulations
illustrates the sensitivity of the luminosity to the positions of the
wind sources.  Luminosities for each of our simulations are reported
in Table~\ref{sims}; for comparison, \citeauthor{BEA03} reported a
2--10 keV luminosity of $(7.6^{+2.6}_{-1.9})\times 10^{31}$ ergs
s$^{-1}$ arcsec$^{-2}$ from the local diffuse emission within
10$^{\prime\prime}$ of Sgr A*.  Results from both simulations fall
within the error bars reported by \citeauthor{BEA03}

Table~\ref{sims} also includes two different estimates for the
temperature of the plasma within 10$^{\prime\prime}$ of Sgr A*.
T$_{mass}$ is an average temperature calculated using the mass of each
particle as a weighting factor, while T$_{lum}$ uses the luminosity of
each particle as a weighting factor.  Since the temperature is
determined primarily by the kinetic energy flux initially carried by
the winds, and since the initial wind velocities in each calculation
are the same, the average temperature for the plasma at the center of
each calculation is essentially the same.

\begin{figure}
\epsscale{1.00}
\plotone{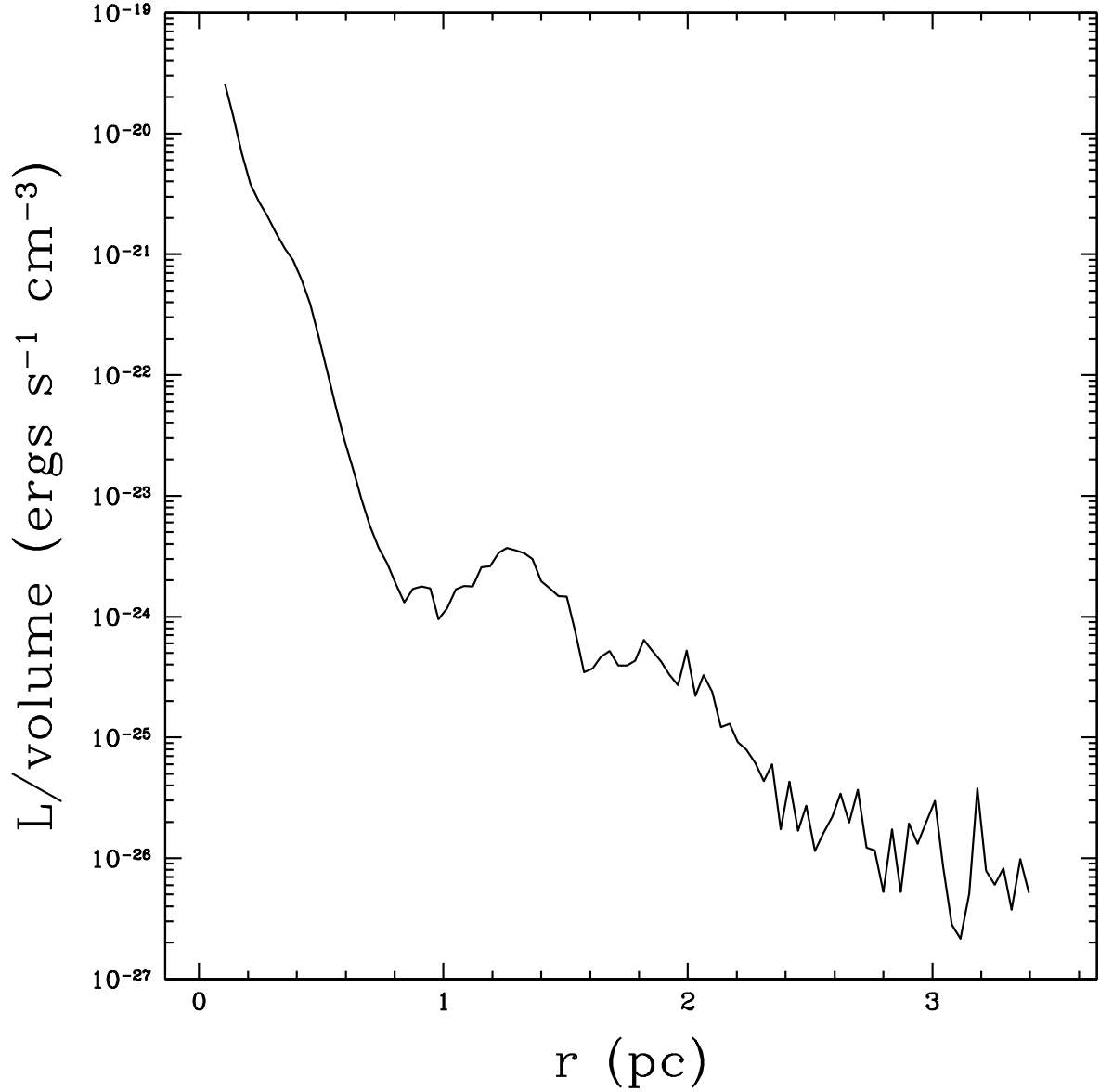}
\caption{2--10 keV X-ray luminosity per volume per 0.035~pc radial 
  ``bin'' versus distance from the central black hole in Simulation 1.
  Collisions between the winds and the inner edge of the CND are
  visible as a bump near 1.2~pc.}
\label{fig-lvsr}
\end{figure}

\begin{figure}
\epsscale{1.00}
\plotone{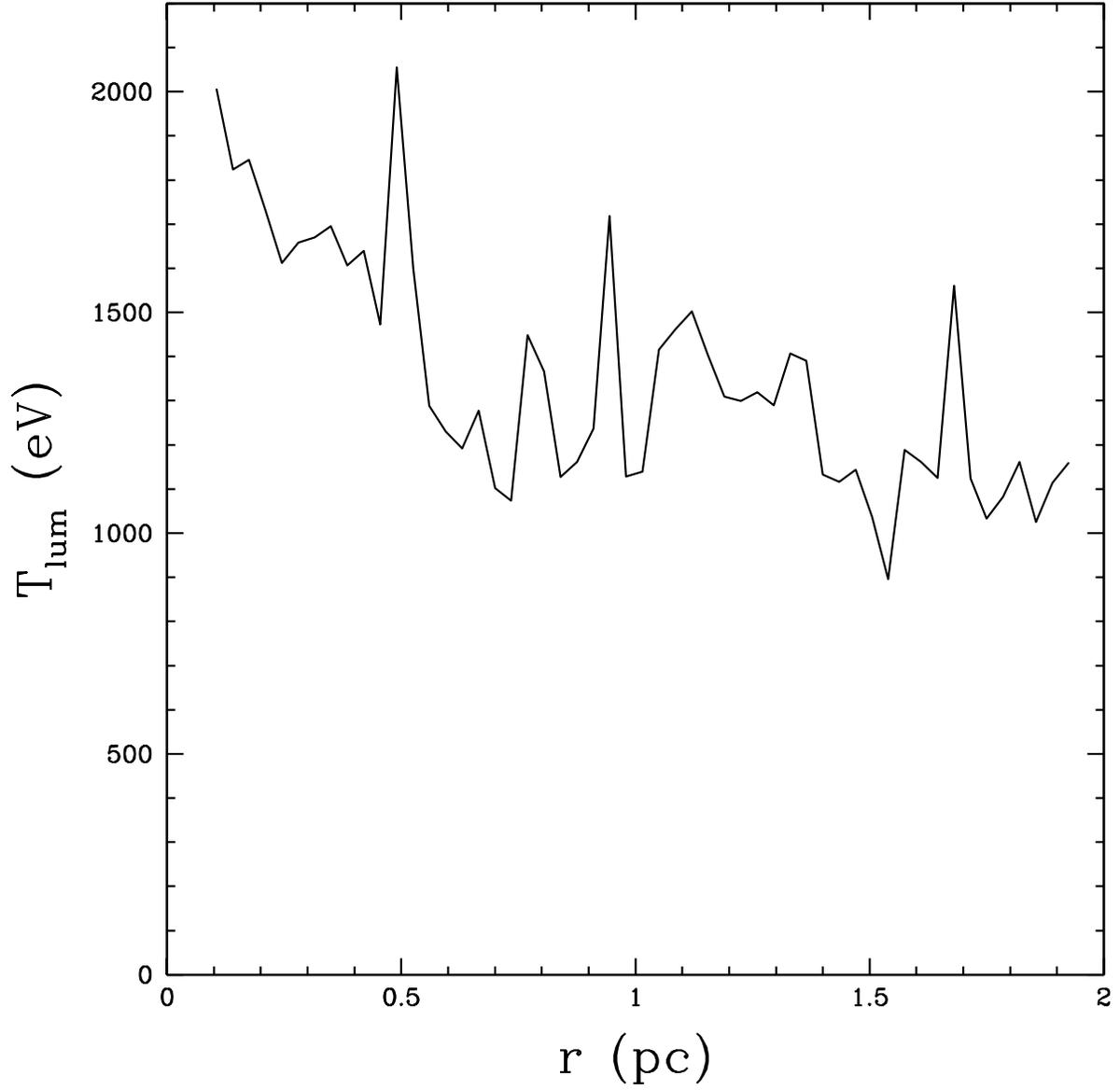}
\caption{Average luminosity-weighted temperature per 0.035~pc radial
  ``bin'' versus distance from the central black hole in Simulation 1.
  Collisions between the winds and the inner edge of the CND are
  visible as a spike near 1~pc.}
\label{fig-tvsr}
\end{figure}

Figures~\ref{fig-lvsr} and \ref{fig-tvsr} show the average luminosity
per volume and average luminosity-weighted temperature as a function
of radius from the center of Simulation 1.  The majority ($77\%$ of
the overall 2--10 keV luminosity) of the 2--10 keV emission comes from
the central 0.4~pc, where it is produced in wind-wind collisions.
There is also noticeable emission ($2.5\%$ of the overall luminosity)
from a high-temperature region centered on 1.2~pc, which is the
location of the inner edge of the CND; emission here is produced when
winds collide with the clumps in our model torus.

\begin{figure}
\epsscale{1.00}
\plotone{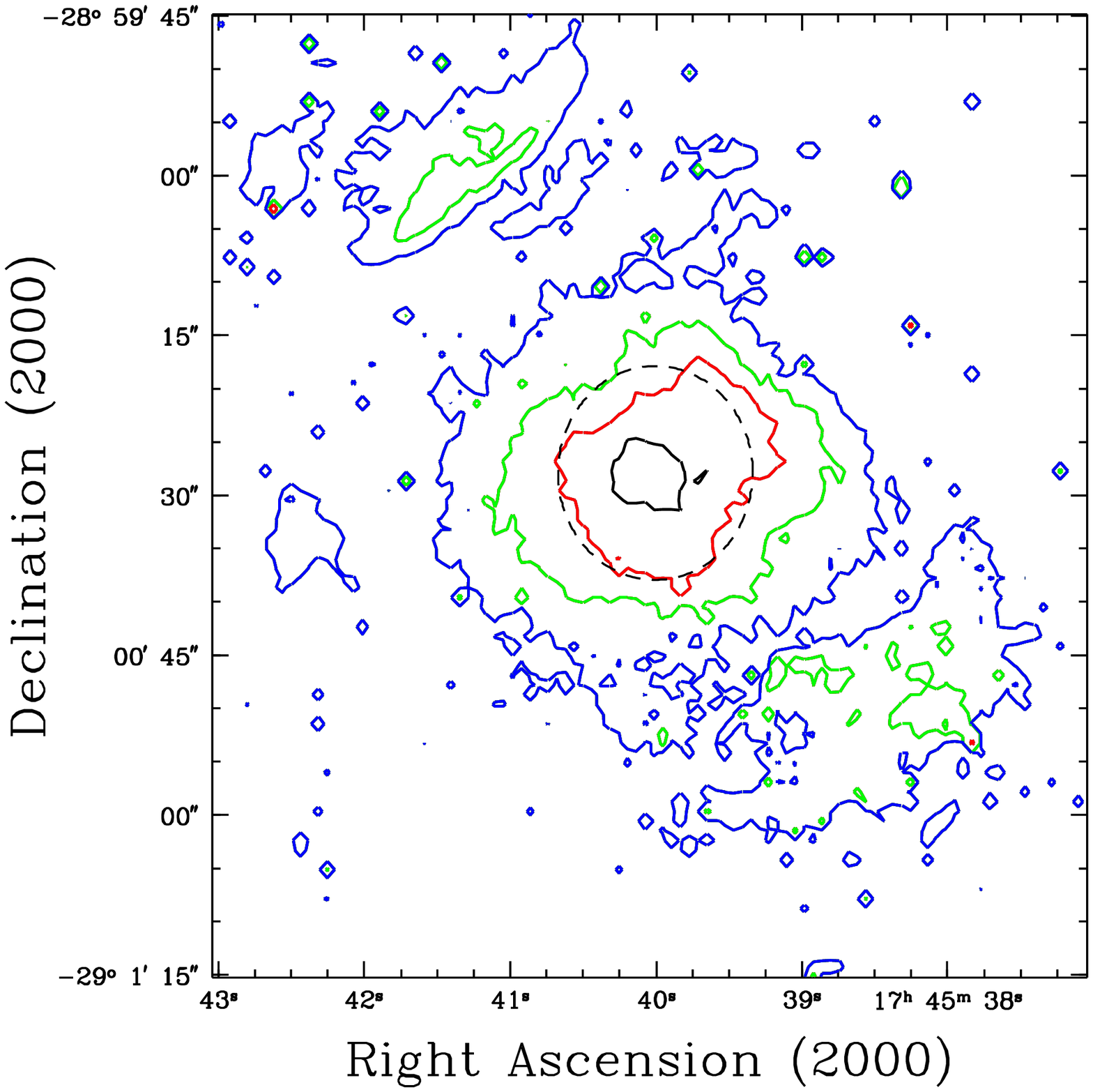}
\caption{Contours of column-integrated 2--10 keV X-ray luminosity per
  ``bin'' from Simulation 1.  There are 100 bins along each axis; each
  bin is 0.9$^{\prime\prime}$ by 0.9$^{\prime\prime}$.  In order from
  black to red to green to blue, the luminosities indicated by the
  contours are $10^{32}$~ergs~s$^{-1}$ per bin,
  $10^{31}$~ergs~s$^{-1}$ per bin, $10^{30}$~ergs~s$^{-1}$ per bin,
  and $1.5\times 10^{29}$~ergs~s$^{-1}$ per bin.  The dashed circle
  indicates the extent of the ``local'' region around Sgr A*; it has a
  radius of $10^{\prime\prime}$ and is centered on Sgr A*.}
\label{contour-orig}
\end{figure}

\begin{figure}
\epsscale{1.00}
\plotone{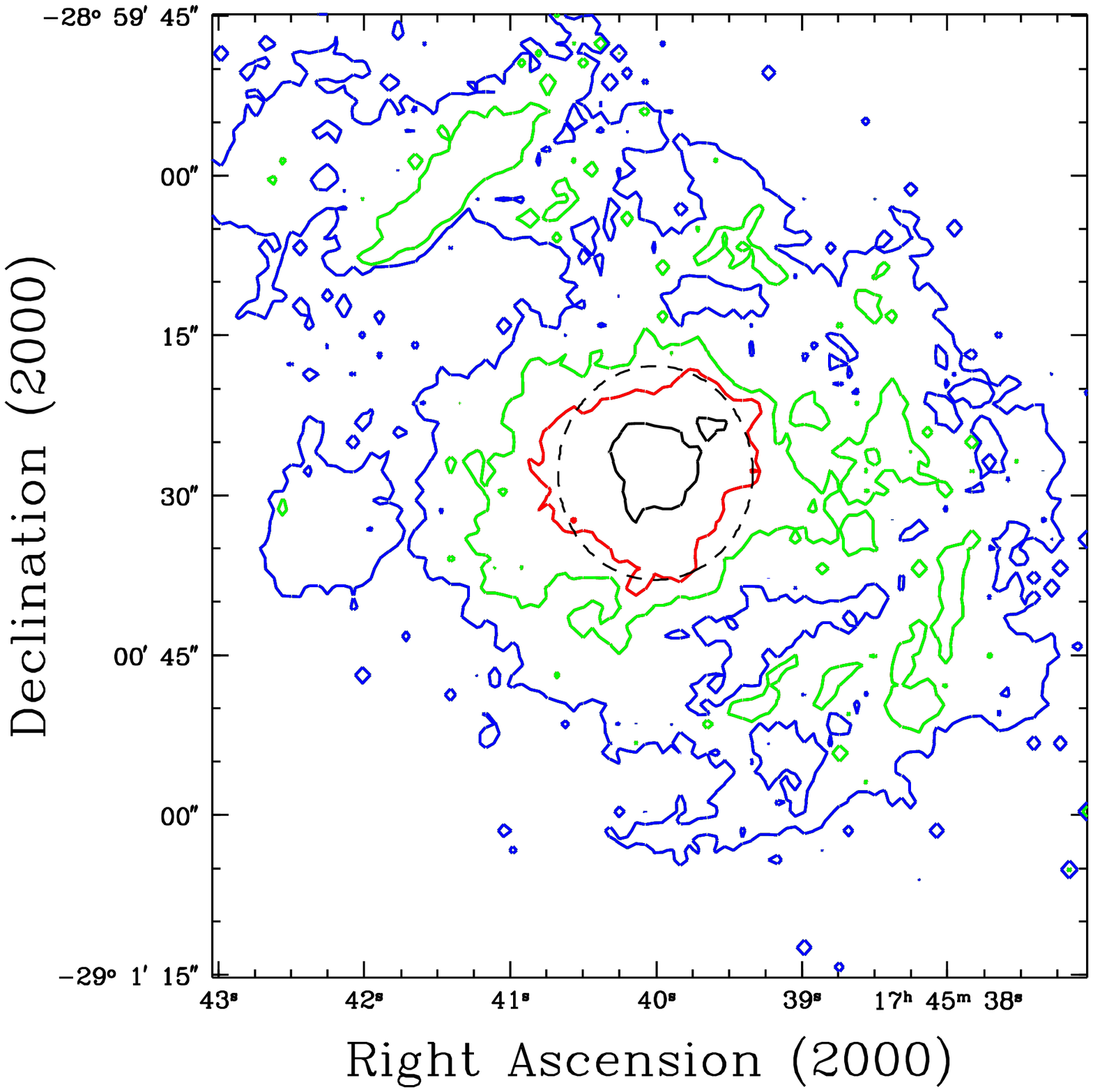}
\caption{Contours of column-integrated 2--10 keV X-ray luminosity per
  ``bin'' from Simulation 2.  There are 100 bins along each axis; each
  bin is 0.9$^{\prime\prime}$ by 0.9$^{\prime\prime}$.  In order from
  black to red to green to blue, the luminosities indicated by the
  contours are $10^{32}$~ergs~s$^{-1}$ per bin,
  $10^{31}$~ergs~s$^{-1}$ per bin, $10^{30}$~ergs~s$^{-1}$ per bin,
  and $1.5\times 10^{29}$~ergs~s$^{-1}$ per bin.  The dashed circle
  indicates the extent of the ``local'' region around Sgr A*; it has a
  radius of $10^{\prime\prime}$ and is centered on Sgr A*.}
\label{contour-mind}
\end{figure}

Figures \ref{contour-orig} and \ref{contour-mind} show contours of
2--10 keV luminosity per arcsec$^{2}$ from Simulations 1 and 2,
respectively, $10,000$ years after the winds were turned on.  The
dotted circle overlaid on each figure shows the size of the
10$^{\prime\prime}$ central region.  The zones of relatively high
luminosity at the north-east and south-west corners of each figure,
and along the northwest side of the Simulation 2 contours, result from
collisions of the outflowing wind with the inner edge of the CND.

The gas distribution in a multiple-wind source environment like that
modeled here is distinctly different from that of a uniform flow past
a central accretor \citep{CM97}.  Unlike the latter, the former does
not produce a large-scale bow shock, and therefore the environmental
impact of the gravitational focusing by the central dark mass has
significantly less order in this case.  We shall defer a more
extensive discussion of this point to a later publication, in which we
report the results of a multiple-wind source simulation for the
accretion of shocked gas onto the central, massive black hole.

\section{Discussion}

Using only the latest observed stellar positions and inferred mass
loss rates and wind velocities, we have produced self-consistent X-ray
luminosity maps of the Galactic center.  It appears that the diffuse
X-ray luminosity and temperature within $10^{\prime\prime}$ of Sgr A*
can be explained entirely by shocked winds; the luminosities from both
of our simulations fall within the error bars associated with the
diffuse 2--10 keV X-ray luminosity measured by {\it Chandra}.  This
seems rather significant in view of the fact that, were the diffuse
X-rays surrounding Sgr A* being produced by another mechanism, we
would need to lower the overall stellar mass-loss rate or wind
velocity by a factor of 2 or more in order to render the emissivity
produced by wind-wind collisions unobservable.  The luminosities that
we calculate are fairly insensitive to the placement of the wind
sources along the line of sight to the Galactic Center; the 2--10 keV
luminosity per arcsec$^2$ differs by only $\sim 15\%$ between
Simulation 1---in which sources are randomly distributed in $z$---and
Simulation 2---in which sources are compressed toward Sgr A*
perpendicular to the plane of the sky.

Although the wind mass-loss rate is $3\times 10^{-3}\;
M_\odot$~yr$^{-1}$, the average accretion rate though our inner
boundary is only $4.25\times 10^{-4}\; M_\odot$~yr$^{-1}$ for Simulation
1 and $4.59\times 10^{-4}\; M_\odot$~yr$^{-1}$ for Simulation 2.  Even
so, this is much higher than what one would expect from the X-ray
emission at this inner boundary.  Since we do not model this
accretion, we have neglected a number of effects from magnetic fields
to radiation pressure.  We postpone this detailed analysis to a future
paper.

One possible explanation for the relatively low luminosity of Sgr A*
has been that the supernova that created Sgr A East swept most of the
gas out of the environment surrounding Sgr A*.  However, our
calculations show that the stellar winds within 3 parsecs of the
Galactic center bring the environment near the black hole back to a
steady state within $\sim$~4000 years.  In addition, the gas
temperature and X-ray luminosity calculated from our results add
weight to \citeauthor{BEA03}'s hypothesis that the X-ray emitting gas
surrounding Sgr A* really is a different component than that at the
center of Sgr A East, which has a temperature closer to 2 keV.

Collision of the wind with the inner edge of the CND may produce the
western arm of the minispiral.  Our simple model for the CND produces
clear emission at the location of the inner edge of the molecular
ring, although we do not reproduce the entire complex morphology of
the X-ray emitting gas seen by {\it Chandra}.

The excellent agreement between our calculated X-ray luminosities and
the value of the luminosity measured by {\it Chandra} solidifies our
understanding of the gas dynamics in the Galactic center.  Future
calculations will move the inner boundary of the simulation inward and
explore the dynamics of the shocked gas as it falls toward the central
black hole.

{\bf Acknowledgments}
This research was partially supported by NASA under grants NAG5-8239
and NAG5-9205, and has made use of NASA's Astrophysics Data System
Abstract Service.  This work was also funded under the auspices of the
U.S. Dept. of Energy, and supported by its contract W-7405-ENG-36 to
Los Alamos National Laboratory and by a DOE SciDAC grant number
DE-FC02-01ER41176.  FM is grateful to the University of Melbourne for
its support (through a Sir Thomas Lyle Fellowship and a Miegunyah
Fellowship).  The authors also thank the anonymous referee for helpful
corrections and comments.  The simulations were conducted on the Space
Simulator at Los Alamos National Laboratory.

{}

\newpage
\begin{deluxetable}{lccccc}
\tablewidth{0pt}
\tablecaption{Simulation Properties\label{sims}}
\tablehead{
  \colhead{Simulation}
& \colhead{z positions}
& \colhead{10$^{\prime\prime}$ L}
& \colhead{T$_{mass}$}
& \colhead{T$_{lum}$} \\

& 
& \colhead{(ergs s$^{-1}$ arcsec$^{-2}$)}
& \colhead{(keV)}
& \colhead{(keV)}
}
\startdata

S1 & z$_1$ & $6.45\times 10^{31}$ & 1.72 & 0.25 \\
S2 & z$_2$ & $7.50\times 10^{31}$ & 1.76 & 0.25 \\

\enddata
\end{deluxetable}
\newpage

\begin{deluxetable}{lcccccc}
\tablewidth{0pt}
\tablecaption{Parameters for Galactic Center Wind Sources\label{srcs}}
\tablehead{
  \colhead{Star}
& \colhead{x\tablenotemark{a}}
& \colhead{y\tablenotemark{a}}
& \colhead{z$_{1}$\tablenotemark{a}}
& \colhead{z$_{2}$\tablenotemark{a}}
& \colhead{v}
& \colhead{\mdot} \\

& \colhead{(arcsec)}
& \colhead{(arcsec)}
& \colhead{(arcsec)}
& \colhead{(arcsec)}
& \colhead{(km s$^{-1}$)}
& \colhead{(${10}^{-5}\;M_\odot$ yr$^{-1}$)}
}
\startdata

IRS 16NE                   &   \phn$-$2.6 &  \phn\phs0.8 & \phn\phs2.2 &  \phn\phs6.8 & \phn\phm{,}550 & \phn9.5 \\
IRS 16NW                   &  \phn\phs0.2 &  \phn\phs1.0 &  \phn$-$8.3 &   \phn$-$5.5 & \phn\phm{,}750 & \phn5.3 \\
IRS 16C                    &   \phn$-$1.0 &  \phn\phs0.2 & \phn\phs4.5 &  \phn\phs2.1 & \phn\phm{,}650 &    10.5 \\
IRS 16SW                   &   \phn$-$0.6 &   \phn$-$1.3 &  \phn$-$2.5 &   \phn$-$1.2 & \phn\phm{,}650 &    15.5 \\
IRS 13E1                   &  \phn\phs3.4 &   \phn$-$1.7 &  \phn$-$0.3 &  \phn\phs1.3 &          1,000 &    79.1 \\
IRS 7W                     &  \phn\phs4.1 &  \phn\phs4.8 &  \phn$-$5.5 &   \phn$-$2.8 &          1,000 &    20.7 \\
AF                         &  \phn\phs7.3 &   \phn$-$6.7 & \phn\phs6.2 &   \phn$-$1.2 & \phn\phm{,}700 & \phn8.7 \\
IRS 15SW                   &  \phn\phs1.5 &     \phs10.1 & \phn\phs8.7 &  \phn\phs0.3 & \phn\phm{,}700 &    16.5 \\
IRS 15NE                   &   \phn$-$1.6 &     \phs11.4 & \phn\phs0.7 &   \phn$-$1.1 & \phn\phm{,}750 &    18.0 \\
IRS 29N\tablenotemark{b}   &  \phn\phs1.6 &  \phn\phs1.4 & \phn\phs8.3 &  \phn\phs3.2 & \phn\phm{,}750 & \phn7.3 \\
IRS 33E\tablenotemark{b}   &  \phn\phs0.0 &   \phn$-$3.0 & \phn\phs0.6 &  \phn\phs6.0 & \phn\phm{,}750 & \phn7.3 \\
IRS 34W\tablenotemark{b}   &  \phn\phs3.9 &  \phn\phs1.6 & \phn\phs4.0 &   \phn$-$4.8 & \phn\phm{,}750 & \phn7.3 \\
IRS 1W\tablenotemark{b}    &   \phn$-$5.3 &  \phn\phs0.3 &  \phn$-$0.2 &   \phn$-$4.5 & \phn\phm{,}750 & \phn7.3 \\
IRS 9NW\tablenotemark{b}   &   \phn$-$2.5 &   \phn$-$6.2 &  \phn$-$3.5 &   \phn$-$4.1 & \phn\phm{,}750 & \phn7.3 \\
IRS 6W\tablenotemark{b}    &  \phn\phs8.1 &  \phn\phs1.6 & \phn\phs3.1 &   \phn$-$0.4 & \phn\phm{,}750 & \phn7.3 \\
AF NW\tablenotemark{b}     &  \phn\phs8.3 &   \phn$-$3.1 &  \phn$-$0.1 &   \phn$-$2.4 & \phn\phm{,}750 & \phn7.3 \\
BLUM\tablenotemark{b}      &  \phn\phs9.2 &   \phn$-$5.0 &  \phn$-$4.1 &  \phn\phs0.2 & \phn\phm{,}750 & \phn7.3 \\
IRS 9S\tablenotemark{b}    &   \phn$-$5.5 &   \phn$-$9.2 &  \phn$-$5.9 &   \phn$-$0.3 & \phn\phm{,}750 & \phn7.3 \\
Unnamed 1\tablenotemark{b} &  \phn\phs1.3 &   \phn$-$0.6 &  \phn$-$5.4 &  \phn\phs5.5 & \phn\phm{,}750 & \phn7.3 \\
IRS 16SE\tablenotemark{b}  &   \phn$-$1.4 &   \phn$-$1.4 &  \phn$-$8.1 &   \phn$-$5.7 & \phn\phm{,}750 & \phn7.3 \\
IRS 29NE\tablenotemark{b}  &  \phn\phs1.1 &  \phn\phs1.8 & \phn\phs3.1 &   \phn$-$3.1 & \phn\phm{,}750 & \phn7.3 \\
IRS 7SE\tablenotemark{b}   &   \phn$-$2.7 &  \phn\phs3.0 & \phn\phs2.3 &   \phn$-$5.4 & \phn\phm{,}750 & \phn7.3 \\
Unnamed 2\tablenotemark{b} &  \phn\phs3.8 &   \phn$-$4.2 &  \phn$-$8.5 &  \phn\phs4.5 & \phn\phm{,}750 & \phn7.3 \\
IRS 7E\tablenotemark{b}    &   \phn$-$4.2 &  \phn\phs4.9 & \phn\phs8.6 &  \phn\phs1.3 & \phn\phm{,}750 & \phn7.3 \\
AF NWW\tablenotemark{b}    &     \phs10.2 &   \phn$-$2.7 &  \phn$-$1.9 &  \phn\phs3.9 & \phn\phm{,}750 & \phn7.3 \\

\enddata
\tablenotetext{a}{Relative to Sgr A* in l-b coordinates where $+$x
is west and $+$y is north of Sgr A*}
\tablenotetext{b}{Wind velocity and mass loss rate fixed (see text)}
\end{deluxetable}
\newpage

\newpage

\newpage

\newpage

\newpage

\newpage

\newpage

\newpage

\newpage

\end{document}